\documentclass[letterpaper]{article} 
\usepackage{aaai2026}  
\usepackage{times}  
\usepackage{helvet}  
\usepackage{courier}  
\usepackage[hyphens]{url}  
\usepackage{graphicx} 
\usepackage{natbib}  
\usepackage{caption} 
\usepackage{algorithm}
\usepackage{algorithmic}

\usepackage{newfloat}
\usepackage{listings}
\DeclareCaptionStyle{ruled}{labelfont=normalfont,labelsep=colon,strut=off} 
\floatstyle{ruled}
\newfloat{listing}{tb}{lst}{}
\floatname{listing}{Listing}
\usepackage{pifont} 
\usepackage{booktabs}
\usepackage{multirow}
\usepackage{amsmath,amssymb,amsfonts}
\usepackage{enumitem}
\usepackage{makecell}
\usepackage{placeins}

\title{Virtual Multiplex Staining for Histological Images Using a Marker-wise Conditioned Diffusion Model} 
\author{
    Hyun-Jic Oh\textsuperscript{\rm 1},
    Junsik Kim\textsuperscript{\rm 2},
    Zhiyi Shi\textsuperscript{\rm 2},
    Yichen Wu\textsuperscript{\rm 2},
    Yu-An Chen\textsuperscript{\rm 3},
    Peter K Sorger\textsuperscript{\rm 3},
    \\
    Hanspeter Pfister\textsuperscript{\rm 2},
    Won-Ki Jeong\textsuperscript{\rm 1}\thanks{Corresponding author: wkjeong@korea.ac.kr} 
}
\affiliations{
    \textsuperscript{\rm 1} Korea University \\ 
    \textsuperscript{\rm 2} Harvard University \\ 
    \textsuperscript{\rm 3} Harvard Medical School, Harvard University \\ 
    
    \{hyunjic0127,wkjeong\}@korea.ac.kr, 
    mibastro@gmail.com, 
    zhiyis@seas.harvard.edu,
    wuyichen.am97@gmail.com,
    pfister@seas.harvard.edu,
    yu-an\_chen@hms.harvard.edu,
    peter\_sorger@hms.harvard.edu
    \\
}

\usepackage{bibentry}

\begin{document}

\maketitle

\begin{abstract}
Multiplex imaging is revolutionizing pathology by enabling the simultaneous visualization of multiple biomarkers within tissue samples, providing molecular-level insights that traditional hematoxylin and eosin (H\&E) staining cannot provide.
However, the complexity and cost of multiplex data acquisition have hindered its widespread adoption. 
Additionally, most existing large repositories of H\&E images lack corresponding multiplex images, limiting opportunities for multimodal analysis.
To address these challenges, we leverage recent advances in latent diffusion models (LDMs), which excel at modeling complex data distributions by utilizing their powerful priors for fine-tuning to a target domain. 
In this paper, we introduce a novel framework for virtual multiplex staining that utilizes pretrained LDM parameters to generate multiplex images from H\&E images using a conditional 
diffusion model. 
Our approach enables marker-by-marker generation by conditioning the diffusion model on each marker, while sharing the same architecture across all markers. 
To tackle the challenge of varying pixel value distributions across different marker stains and to improve inference speed, we fine-tune the model for single-step sampling, enhancing both color contrast fidelity and inference efficiency through pixel-level loss functions. 
We validate our framework on two publicly available datasets, 
notably demonstrating its effectiveness in generating up to 18 different marker types with improved accuracy, a substantial increase over the 2-3 marker types achieved in previous approaches. This validation highlights the potential of our framework, pioneering virtual multiplex staining.
%
Finally, this paper bridges the gap between H\&E and multiplex imaging, potentially enabling retrospective studies and large-scale analyses of existing H\&E image repositories.
\end{abstract}

\begin{figure}[ht]
    \centering
    \includegraphics[width=0.95\columnwidth]{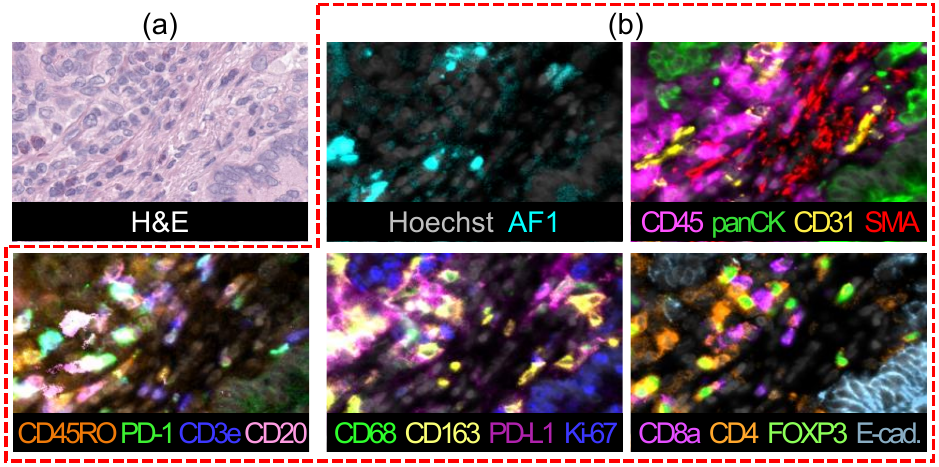}
    \caption{
    Sample from Orion-CRC~\cite{lin2023high}: (a) H\&E staining and (b) multiplex immunofluorescence (mIF) of the same region.
    Each color in mIF corresponds to a different marker.
    }
    \label{fig:intro}
\end{figure}

\section{Introduction}
\label{sec:introduction}

Histopathological analysis is crucial for disease diagnosis and biomedical research, enabling the identification of pathological changes and guiding treatment decisions. 
%
Among available techniques, hematoxylin and eosin (H\&E) staining has long been the gold standard, often complemented by immunohistochemistry (IHC). 
Despite their widespread use, both H\&E and IHC have limitations in capturing the complex tissue microenvironment, especially in advanced cancer studies~\cite{wharton2021tissue}.
To address these limitations, multiplex immunofluorescence (mIF) and multiplex immunohistochemistry (mIHC) imaging have emerged, enabling visualization of multiple biomarkers within a single tissue sample and providing more comprehensive insights into the tissue microenvironment (see Fig.~\ref{fig:intro})~\cite{sood2016multiplexed,tan2020overview,lin2023high}. 
%
However, widespread adoption of multiplex imaging is hindered by complex imaging protocols and high staining costs.

To overcome these challenges, computational approaches known as \textit{virtual staining}, which generate various staining images from H\&E stained tissue images, have gained much attention~\cite{latonen2024virtual}.
Most prior work applied deep generative models to unpaired H\&E-to-IHC tasks, focusing on spatial alignment~\cite{li2023adaptive,chen2024pathological,wang2025oda}; other studies explored multiplex marker generation, such as H\&E-to-IF~\cite{burlingame2020shift} and H\&E-to-mIHC~\cite{pati2024accelerating}.
%
However, these approaches generally require a separate model per marker, limiting scalability and increasing computational cost, which hinders efficient handling of large marker sets.
Moreover, their separate training manner focuses only on individual markers, missing valuable cross-channel relationships and overlooking the potential benefits of knowledge sharing through joint training.
Even recent methods like VIMs~\cite{dubey2024vims} and HEMIT~\cite{bian2024hemit}, which generate up to two or three markers, still lack the scalability needed for practical multiplex imaging applications.

Based on these observations, we propose a novel framework for virtual multiplex staining that utilizes a conditional diffusion model~\cite{ho2020denoising,dhariwal2021diffusion,rombach2022high} to translate H\&E images into multiplex marker images.
By designing a marker-wise generation approach, we aim to generate multiple markers within a single shared model architecture, enabling knowledge sharing during training and reducing the significant computational overhead. Specifically, to optimize both the efficiency of the generation process and the quality of the generated results, we adopt a two-stage training approach, each addressing a core challenge: 

\noindent{\textbf{(1) Multi-target generation.}}
Unlike previous methods training separate models for each marker or lacking scalability for the large number of marker types, we use a marker-wise conditioning strategy to generate multiple markers simultaneously using a single model.
By conditioning H\&E images with different marker-type embeddings using latent diffusion models~\cite{rombach2022high}, 
we enable the model to effectively distinguish among a large set of markers and generate high-quality multiplex staining from H\&E images, overcoming the limitations of conventional text-based conditioning when the number of marker types increases.
Therefore, the training strategy reduces computational costs and enables potential knowledge sharing.

\noindent{\textbf{(2) Color contrast fidelity.}} 
Due to variations in pixel value ranges and contrasts across marker images, simply applying diffusion models tends to be influenced by the training data bias, where a substantial portion of images contain dark background pixels.
To address this issue, we incorporate a second training stage that fine-tunes the model using pixel-level loss functions to enhance marker-specific generation capabilities, addressing color accuracy. 
This stage also optimizes the model for fast, single-step inference, enabling practical deployment in pathology where timely analysis impacts clinical workflows and large-scale research studies.

Through the proposed two-stage training framework, we efficiently generate high-quality multiplex marker images from H\&E staining, preserving spatial and structural context and improving color fidelity, thereby promoting the widespread adoption of multiplex imaging.
To the best of our knowledge, this is the first work to demonstrate virtual staining of up to 18 distinct marker images from a single H\&E image using a conditional diffusion model.
To summarize, our contributions are as follows:
\begin{itemize}[leftmargin=3mm, itemsep=0mm, topsep=-0.5 mm]
    \item We propose a novel two-stage training framework for virtual multiplex staining from H\&E tissue sections, leveraging a specially designed conditional diffusion model embedded with marker-type embeddings to effectively handle a large number of marker types.
    \item We address the issue of color fidelity during the second-stage fine-tuning phase by integrating pixel-level loss functions, while also opting to optimize the single-step sampling of the conditional diffusion model for faster inference.
    \item We validate our framework on two publicly available datasets, demonstrating its effectiveness in generating up to 18 distinct marker types with superior accuracy.
\end{itemize}

\section{Related Work}
\label{sec:rel_work}

\begin{figure*}
    \centering
    \includegraphics[width=0.9\linewidth]{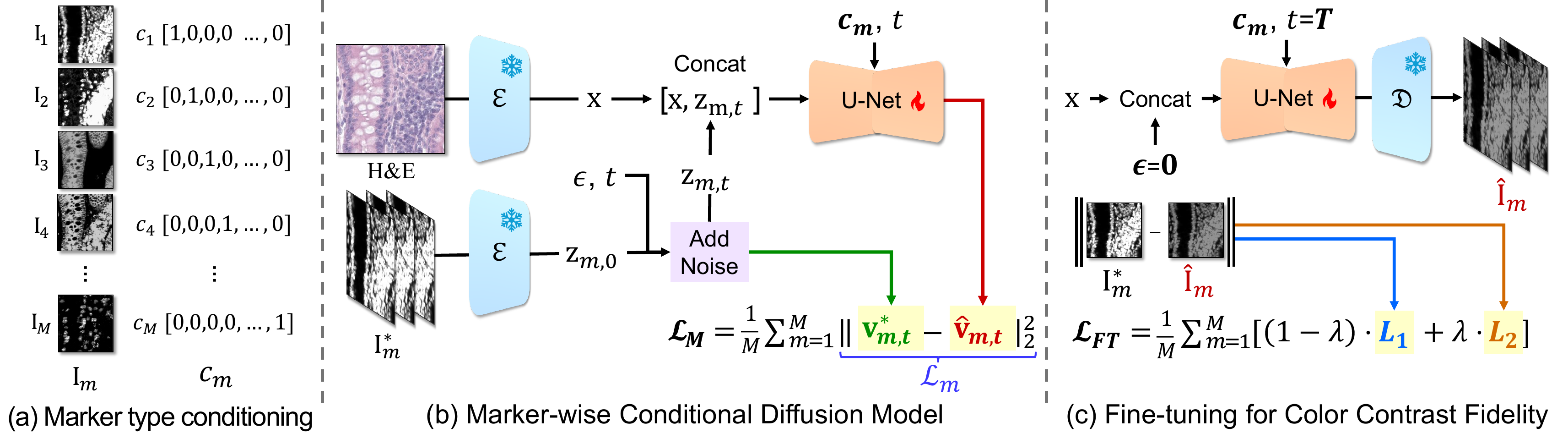}
    \caption{
    Overview of the proposed two-stage training framework for virtual multiplex staining.
    (a) Marker type conditioning using the one-hot vector $c_m$ for marker $m$.
    (b) Marker-wise conditional diffusion model that generates multiplex marker images from an H\&E input with marker-specific one-hot conditioning.
    (c) Pixel-level fine-tuning stage to improve color fidelity, enabling single-step inference. 
    }
    \label{fig:train}
\end{figure*}

\subsection{Multiplex Imaging for Pathology}

H\&E staining is the gold standard for tissue examination in pathology, with IHC staining complementing morphological assessment by providing additional molecular information.
Previous virtual staining studies have mainly focused on translating H\&E to IHC images in unpaired settings to address spatial alignment challenges inherent between these stains~\cite{li2023adaptive,chen2024pathological,wang2025oda}.
While these methods have advanced unpaired translation, their objectives and data settings differ from our focus on paired H\&E-to-multiplex virtual staining.
Moreover, both H\&E and IHC stainings are limited in capturing the complex tissue microenvironment, particularly in advanced cancers~\cite{wharton2021tissue}.
This has led to the increased interest in multiplex imaging techniques, which simultaneously visualize multiple biomarkers within a single tissue section—enhancing cellular, molecular, and spatial understanding of pathology~\cite{tan2020overview,wharton2021tissue,lin2023high}.
Despite its analytic advantages, the widespread adoption of multiplex imaging is hindered by resource-intensive protocols and limited data availability. 

To address these challenges, research has shifted toward deep learning-based virtual staining methods that generate multiplex images from H\&E sections~\cite{burlingame2020shift,pati2024accelerating,bian2024hemit}.
Early efforts relied on pix2pix~\cite{isola2017image}, a conditional GAN used as a baseline for paired image-to-image translation from H\&E to specific immunomarkers, which required a separate model per marker, thus limiting scalability as the number of biomarkers increases~\cite{burlingame2020shift}.
Recently, HEMIT~\cite{bian2024hemit} combined residual CNNs with Swin Transformers to jointly generate up to three mIHC markers, partially capturing cross-marker relationships; however, it has yet to demonstrate scalability to larger marker panels.
Meanwhile, VIMs~\cite{dubey2024vims} used an unpaired, diffusion-based approach to generate two IHC markers with performance comparable to pix2pix, but it requires expert-designed text prompts for marker control, which poses barriers when scaling to more markers or deploying in automated pipelines.

To sum up, existing approaches remain limited by either per-marker model requirements, restricted marker panel size, or dependencies on manual prompt engineering. Efficient joint modeling and scalability to high marker counts, as well as comprehensive knowledge sharing, thus remain open challenges.
Our approach directly addresses these limitations by introducing a conditional diffusion model-based framework capable of generating multiple marker types, paving the way for more scalable and practical virtual multiplex imaging from H\&E staining.

\subsection{Conditional Diffusion Models}
Diffusion models~\cite{sohl2015deep,ho2020denoising} have emerged as a powerful generative framework, achieving remarkable fidelity in image synthesis via iterative denoising.
Of particular interest, Latent Diffusion Models (LDMs)~\cite{rombach2022high} leverage a Variational Autoencoder (VAE) trained on large-scale datasets to operate in a compressed latent space, substantially improving computational efficiency and providing strong priors for rapid adaptation in diverse domains.

Recent studies have demonstrated the versatility of diffusion models for a wide range of image generation.
Conditional diffusion models allow fine-grained control over image generation by leveraging diverse conditioning inputs, including text~\cite{saharia2022photorealistic}, semantic maps~\cite{zhang2023adding}, or images~\cite{tumanyan2023plug}. Notably, these models have shown strong performance in dense prediction tasks like semantic segmentation, depth estimation, and structured image-to-image translation by effectively integrating conditioning signals into the diffusion process~\cite{zhao2023unleashing,ke2024repurposing,fu2025geowizard}.
Building on these developments, approaches like Marigold~\cite{ke2024repurposing} keep the VAE prior from pre-trained LDMs and fine-tune only the denoising U-Net for specific tasks, such as depth prediction, achieving efficient adaptation with reduced training complexity.
Similarly, Geowizard~\cite{fu2025geowizard} introduces a class-wise one-hot conditioning strategy to differentiate multiple geometric and domain classes, which is beneficial when scaling to many output targets conditioned on the same image prompt.

Despite these advances, existing conditional diffusion models remain under-explored for multiplexed pathology images, which pose distinct challenges such as channel scalability, color fidelity, and computational efficiency. 
%
Our work addresses these challenges by leveraging robust priors from pre-trained LDMs and introducing marker-wise conditioning for scalable, high-fidelity multiplex virtual staining.

\section{Method}
\label{sec:method}

In this section, we introduce our two-stage training framework for virtual multiplex staining in H\&E images using a conditional diffusion model for marker-specific generation.
%
The two stages address key challenges of multiplexed image generation and color contrast fidelity using marker specific conditioning as illustrated in Fig.~\ref{fig:train}(a), with further details provided in Sec.~\ref{sec:diffusion} for (b) a marker-wise conditional diffusion model and in Sec.~\ref{sec:finetuning} for (c) a fine-tuning process to enhance color fidelity. 

\subsection{Marker-wise Conditional Diffusion Model}
\label{sec:diffusion}
For the multiplex image generation, we build our virtual staining model on the pre-trained stable diffusion (SD) backbone~\cite{rombach2022high}, leveraging its LDM architecture. Specifically, we formulate the virtual staining as a conditional diffusion process that maps H\&E images to generate multiplex marker images.

\subsubsection{Image encoding.}
We replicate a single-channel marker image $\mathbf{I}_m$ into three channels to match the input shape of the pre-trained SD VAE $\mathcal{E}$, where $m \in \{1,2,...,M\}$ denotes a marker type.
During training, we encode the ground-truth marker image through the VAE encoder to obtain the clean latent representation $\mathbf{z}_{m,0}=\mathcal{E}(\mathbf{I}_m)$.
Correspondingly, the H\&E image is encoded as $\mathbf{x}$. 
%

\subsubsection{Training process.}
Following the DDPM framework~\cite{ho2020denoising}, we define a timestep $t \in \{1, ..., T\}$ and a variance schedule $\{\beta_1, ..., \beta_T\}$ for diffusion process.
Specifically, during the training stage, we corrupt the ground-truth marker latent $\mathbf{z}_{m,0}$ by adding noise at timestep $t$:
\begin{equation}
    \mathbf{z}_{m,t} = \sqrt{\bar{\alpha}_t} \mathbf{z}_{m,0} + \sqrt{ 1 - \bar{\alpha}_t} \epsilon,
\end{equation}
where $\epsilon \sim \mathcal{N}(0,I)$ is Gaussian noise and $\bar{\alpha}_t = \prod_{k=1}^t (1-\beta_k)$. 
Note that $\mathbf{z}_{m,0}$ is only available during training from ground-truth marker images.
This forward process gradually adds noise to the latent representation, allowing the U-Net to learn the denoising operation.

By freezing the SD VAE encoder $\mathcal{E}$ and decoder $\mathcal{D}$ parameters, we train only the conditional diffusion U-Net, which serves as the denoising network $\hat{\mathbf{v}}_{m,t}(\cdot)$. Specifically, we concatenate a latent pair $[\mathbf{x}, \mathbf{z}_{m,t}]$ along the channel dimension at a timestep $t$ as the input. To accommodate this input, we double the U-Net's input channel size by duplicating weights and halving their values~\cite{ke2024repurposing}. 
We train only the U-Net denoising network $\hat{\mathbf{v}}_{m,t} = \hat{\mathbf{v}}_\theta([\mathbf{x}, \mathbf{z}_{m,t}], t)$ using the following loss function:
\begin{equation}
    \mathcal{L}_{m} = ||\mathbf{v}^*_{m,t} - \mathbf{\hat{v}}_{m,t}||^2_2,
    \label{eq:loss_1}
\end{equation}
where $\mathbf{v}^*_{m,t}=\sqrt{\bar{\alpha}_t}\epsilon - \sqrt{(1-\bar{\alpha}_t)}\mathbf{z}_{m,0}$ represents the training target for noise prediction. 
This formulation ensures the model learns to accurately predict the noise added during the forward diffusion process.
%
\subsubsection{Inference procedure.}
During inference, the generation process starts from random noise $\mathbf{z}_{m,T} \sim \mathcal{N}(0,I)$ and iteratively denoises using the learned U-Net. Starting from $t=T$ and stepping down to $t=1$, we sample:
\begin{equation}
    \mathbf{z}_{m,t-1}=\frac{1}{\sqrt{\bar{\alpha}_t}}(\mathbf{z}_{m,t} - \sqrt{1-\bar{\alpha}_t}\hat{\mathbf{v}}_{m,t}([\mathbf{x},\mathbf{z}_{m,t}], t, c_m).
\end{equation}
%
Note that we do not use ground-truth $\mathbf{z}_{m,0}$ during inference. After reaching $t=0$, the final denoised latent $\hat{\mathbf{z}}_{m,0}$ is reconstructed into pixel space via the frozen VAE decoder: $\mathbf{\hat{I}}_m=\mathcal{D}(\hat{\mathbf{z}}_{m,0})$.
The reconstructed image is averaged across channels to produce a single-channel marker image prediction.
%

\subsubsection{Marker-wise conditioning.}
To enable multiplex staining generation, we condition the diffusion U-Net with marker-specific embedding vectors as depicted in Fig.~\ref{fig:train}(a). This approach avoids the computational inefficiency of training separate models for each marker type.
Moreover, as depicted in Fig.~\ref{fig:train}(b), we replicate the H\&E image latent $\textbf{x}$ to the number of marker types $M$, allowing for balanced parameter update across all marker types during training.
Correspondingly, we modify the loss function $\mathcal{L}_m$ as 

\begin{equation}
    \mathcal{L}_{M} =  \frac{1}{M} \sum_{m=1}^M \mathcal{L}_m=\frac{1}{M} \sum_{m=1}^M ||\mathbf{v}^*_{m,t} - \mathbf{\hat{v}}_{m,t}||^2_2,
    \label{eq:multiplex_loss}
\end{equation}
where $\mathbf{v}^*_{m,t}$ and $\mathbf{\hat{v}}_{m,t} = \hat{\mathbf{v}}_\theta([\mathbf{x}, \mathbf{z}_{m,t}], t, c_m)$ denote the optimization target and prediction for marker type $m$, respectively, with $c_m$ denoting marker-wise conditioning.

Inspired by~\citet{fu2025geowizard}, we implement marker-wise conditioning by representing each marker type as a one-hot vector.
Unlike text conditioning, which can be ambiguous with many marker types, we adopt marker-wise one-hot embedding to provide distinct and scalable signals for multi-marker generation.
The one-hot vector undergoes positional encoding and is then element-wise added to the time embedding to condition the model on the specific marker type $m$.
%
This technique enables the diffusion U-Net to efficiently learn marker-specific features, allowing for multiple marker image generation.
A detailed analysis is provided in Sec.~\ref{sec:results}.
%

\subsection{Fine-tuning for Color Contrast Fidelity}
\label{sec:finetuning}
While our model achieves multiplexed image generation in Sec.~\ref{sec:diffusion}, color contrast fidelity remains a primary challenge due to dataset bias toward dark background regions.
The visual comparison is detailed in Sec.~\ref{sec:results}.
Moreover, efficient processing can also be important for large-scale pathology data, as diffusion models are intrinsically slow due to their iterative process. While approaches such as DDIM~\cite{song2020denoising} help accelerate inference, other techniques like test-time ensembling~\cite{ke2024repurposing} introduce computational overhead for improved performance. 

To address both color fidelity and inference efficiency, we introduce a second, fine-tuning stage as shown in Fig.~\ref{fig:train}(c). 
In this stage, inspired by recent latent diffusion advances~\cite{garcia2024fine}, we enable single-step inference by fixing $t = T$ and replacing the random noise with zero noise ($\epsilon=0$) 
, optimizing the model directly in pixel space.
This modification allows for rapid, deterministic image generation and, crucially, supports the direct application of task-specific pixel-level supervision for improved color contrast. 
Consistent with previous training phases, we update only the parameters of the diffusion U-Net, while keeping the VAE encoder and decoder fixed.

We employ a pixel-level fine-tuning loss $\mathcal{L}_{FT}$ as follows:
\begin{equation}
    \mathcal{L}_{FT} = \frac{1} M \sum_{m=1}^M [ (1-\lambda)\|\mathbf{I}^*_m - \mathbf{\hat{I}}_m\|_1 + \lambda \|\mathbf{I}^*_m - \mathbf{\hat{I}}_m\|_2^2 ],
    \label{eq:pixel_loss}
\end{equation}
where $\mathbf{\hat{I}}_m$ and $\mathbf{I}^*_m$ denote the prediction and ground truth for marker type $m$, respectively.
And we combine the $L_1$ loss $\|\cdot\|_1$ and $L_2$ loss $\|\cdot\|_2^2$ at the pixel level after passing through the fixed decoder $\mathcal{D}$. 
By directly applying this combined loss at the pixel level rather than the latent space, we significantly enhance the marker-specific color fidelity and overall visual quality. 
Details and further analysis of the hyperparameter $\lambda$ within Eq.~(\ref{eq:pixel_loss}) can be found in Sec.~\ref{sec:results}.

\section{Experiments}
\label{sec:results}

\subsection{Setup}
\subsubsection{Datasets.}
We utilized two public datasets, HEMIT~\cite{bian2024hemit} and Orion-CRC~\cite{lin2023high}, each providing paired H\&E and multiplex-stained images, thus offering a robust benchmark for evaluating virtual multiplex staining techniques.
\textbf{(1) HEMIT} contains H\&E images and three mIHC markers: DAPI (highlighting cell nuclei), pan-cytokeratin (panCK, identifying tumor regions), and CD3 (marking T cells), all crucial for analyzing the tumor microenvironment and immune activity.
The dataset consists of 5,292 1024$\times$1024 pixel image pairs with 50\% overlap, split into training (3,717 pairs), validation (630 pairs), and test (945 pairs) sets.
%
For rigorous evaluation, we curated 292 image pairs from the test set by excluding spatially overlapping regions and filtering out empty patches that can hinder meaningful quantitative evaluation.
This strategy prevents artificial inflation of metrics (\textit{e.g.}, PSNR values above 60 dB) observed in biologically irrelevant, low-signal regions.
\begin{table*}[!ht]
\centering
\setlength{\tabcolsep}{1.45mm}
{\fontsize{9pt}{9.5pt}\selectfont
\begin{tabular}{lcccccccccccc}
\toprule   
\specialrule{0.1pt}{0pt}{0pt}
& \multicolumn{4}{c}{SSIM} & \multicolumn{4}{c}{R} & \multicolumn{4}{c}{PSNR (dB)} \\
\cmidrule(lr){2-5} \cmidrule(lr){6-9} \cmidrule(lr){10-13}
Method & DAPI & CD3 & panCK & Avg. & DAPI & CD3 & panCK & Avg. & DAPI & CD3 & panCK & Avg. \\

\midrule

pix2pix~\cite{isola2017image}
& 0.750 & 0.879 & 0.572 & 0.734 & 0.901 & 0.381 & 0.586 & 0.623 & 31.53 & 24.82 & 26.29 & 27.55 \\

pix2pixHD~\cite{wang2018high}
& 0.702 & 0.844 & 0.579 & 0.709 & 0.948 & \underline{0.592} & \underline{0.726} & \underline{0.755} & 33.36 & \underline{26.14} & 28.07 & 29.19 \\

HEMIT~\cite{bian2024hemit}
& 0.769 & \underline{0.882} & \underline{0.659} & \underline{0.770} & \underline{0.964} & 0.567 & 0.707 & 0.746 & 32.82 & 25.90 & 27.63 & 28.78 \\

\citet{parmar2024one}
& 0.606 & 0.845 & 0.652 & 0.701 & 0.924 & 0.332 & 0.595 & 0.617 & 32.15 & 24.09 & 26.07 & 27.44 \\

Marigold~\cite{ke2024repurposing}
& \underline{0.801} & 0.711 & 0.546 & 0.686 & 0.956 & 0.581 & 0.711 & 0.750 & \underline{33.91} & 26.04 & \underline{28.13} & \underline{29.36} \\

\textbf{Ours}
& \textbf{0.855} & \textbf{0.889} & \textbf{0.763} & \textbf{0.836} & \textbf{0.972} & \textbf{0.633} & \textbf{0.781} & \textbf{0.795} & \textbf{35.40} & \textbf{26.53} & \textbf{29.86} & \textbf{30.60} \\

\toprule   
\specialrule{0.1pt}{0pt}{0pt}
\end{tabular}
}
\caption{
Quantitative comparison on the HEMIT dataset (\textbf{Bold}: Best, \underline{Underline}: Second best).
}
\label{table:hemit}
\end{table*}
\begin{figure*}[t]
\centering
\includegraphics[width=0.85\linewidth]{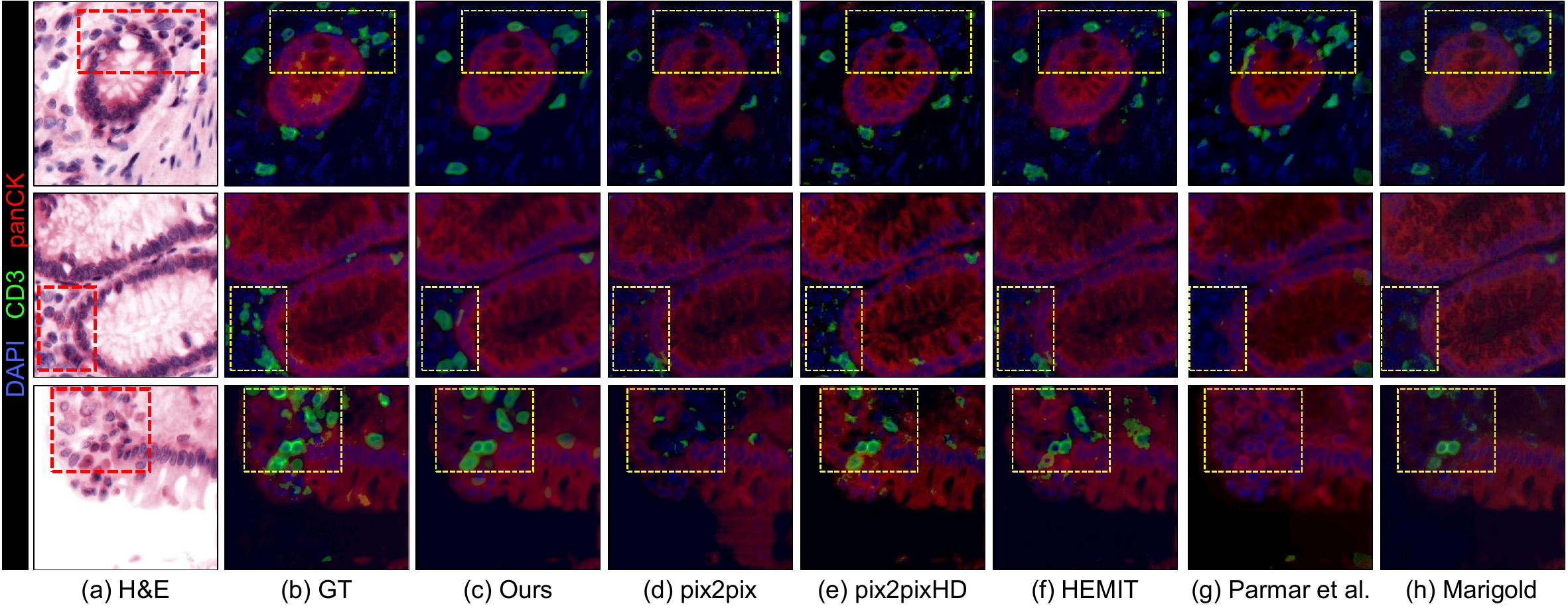}
\caption{
Qualitative comparison on the HEMIT dataset. Each marker type is visualized in distinct colors.
}
\label{fig:qual_hemit}
\end{figure*}
%
%
%
%
%
%
%
%
\textbf{(2) Orion-CRC} comprises 41 colon cancer whole slide images (WSIs), each with 18 IF marker channels paired with H\&E stained sections. 
These include Hoechst (for nuclei), as well as markers capturing immune, epithelial, and cell state features, all informative for spatial tumor characterization.
WSIs (average 81.6G pixels) were cropped into 512$\times$512 patches. After filtering out low-tissue-content patches, the dataset consists of 132,610 training patches (31 WSIs) and 39,625 test patches (10 WSIs).
For quantitative analysis, patches with an SSIM $>0.8$ to an empty array were excluded for each marker to avoid metric inflation due to biologically irrelevant regions.
Due to marker rarity, fewer than 1,000 patches were available for CD31, FOXP3, and CD8a.

\subsubsection{Evaluation metrics.} 
We use the Structural Similarity Index Measure (SSIM), Pearson correlation score (R), and PSNR, following previous works~\cite{burlingame2020shift,bian2024hemit}.

\subsubsection{Implementation details.}
We use SD v2 as the backbone~\cite{rombach2022high}, with patch sizes of 512$\times$512. 
Our models were trained and fine-tuned on four NVIDIA H100 GPUs following the settings of~\citet{ke2024repurposing} and~\citet{garcia2024fine}.
Further details are provided in the supplementary material.

\begin{table*}[t!]
    \setlength{\tabcolsep}{0.68mm}
    {\fontsize{9pt}{9.5pt}\selectfont 
    \centering
    \begin{tabular}{c| l c c c c c c c c c c c c c c c c c c c}
    \toprule
    \specialrule{0.1pt}{0pt}{0pt} 
    \rotatebox{90}{Method} & \; \ \rotatebox{90}{Metric} & \rotatebox{90}{Hoechst} & \rotatebox{90}{AF1} & \rotatebox{90}{CD31} & \rotatebox{90}{CD45} & \rotatebox{90}{CD68} & \rotatebox{90}{CD4} & \rotatebox{90}{FOXP3} & \rotatebox{90}{CD8a} & \rotatebox{90}{CD45RO} & \rotatebox{90}{CD20} & \rotatebox{90}{PD-L1} & \rotatebox{90}{CD3e} & \rotatebox{90}{CD163} & \rotatebox{90}{E-cad.} & \rotatebox{90}{PD-1} & \rotatebox{90}{Ki67} & \rotatebox{90}{panCK} & \rotatebox{90}{SMA} & \rotatebox{90}{Avg.} \\
    \midrule
    \multirow{3}{*}{\makecell{\rotatebox{90}{pix2pix}}} 
    & SSIM & 0.751 & \textbf{0.866} & \underline{0.776} & \underline{0.709} & \underline{0.779} & \underline{0.709} & \textbf{0.638} & 0.654 & 0.739 & \textbf{0.755} & 0.692 & \underline{0.696} & {0.713} & \underline{0.680} & \underline{0.754} & \underline{0.710} & \underline{0.732} & 0.673 & \underline{0.724} \\
    & R   & 0.891 & 0.473 & \textbf{0.096} & \underline{0.427} & \underline{0.173} & \underline{0.097} & \textbf{0.068} & \underline{0.096} & 0.304 & \underline{0.180} & \underline{0.205} & \underline{0.308} & \underline{0.088} & \underline{0.434} & \underline{0.103} & \underline{0.309} & \underline{0.531} & \underline{0.213} & \underline{0.277} \\
    & PSNR & 26.54 & 42.27 & 37.83 & \underline{31.34} & \underline{36.91} & \underline{39.13} & 32.05 & \underline{30.45} & 36.61 & \underline{37.92} & \underline{33.00} & 30.87 & 30.72 & \underline{31.15} & \underline{36.21} & \underline{32.86} & \underline{29.67} & 31.12 & \underline{33.70} \\
    \midrule
    \multirow{3}{*}{\makecell{\rotatebox{90}{HEMIT}}} 
    & SSIM & \underline{0.762} & 0.835 & 0.762 & 0.590 & 0.726 & 0.654 & {0.622} & \textbf{0.715} & \underline{0.752} & 0.578 & \underline{0.726} & 0.682 & \textbf{0.721} & 0.629 & 0.700 & 0.660 & 0.613 & \underline{0.683} & 0.690 \\
    & R    & \underline{0.902} & \textbf{0.533} & {-0.001} & 0.256 & 0.000 & -0.001 & \underline{0.000} & 0.002 & \underline{0.343} & 0.000 & 0.050 & 0.143 & 0.000 & 0.281 & 0.071 & 0.262 & 0.221 & 0.000 & 0.170 \\
    & PSNR & \underline{26.86} & \underline{42.77} & \underline{37.85} & 30.06 & 36.56 & 38.69 & \underline{32.09} & 28.70 & \underline{37.54} & 36.74 & 32.05 & \underline{31.47} & \underline{31.06} & 30.56 & 36.15 & 32.44 & 28.33 & \underline{34.02} & 33.55 \\
    \midrule
    \multirow{3}{*}{\textbf{\rotatebox{90}{Ours}}}
    & SSIM & \textbf{0.779} & \underline{0.852} & \textbf{0.785} & \textbf{0.770} & \textbf{0.815} & \textbf{0.803} & \underline{0.623} & \underline{0.712} & \textbf{0.838} & \underline{0.718} & \textbf{0.753} & \textbf{0.724} & \underline{0.718} & \textbf{0.771} & \textbf{0.764} & \textbf{0.793} & \textbf{0.747} & \textbf{0.761} & \textbf{0.763} \\
    & R   & \textbf{0.912} & \underline{0.510} & \underline{0.027} & \textbf{0.563} & \textbf{0.319} & \textbf{0.340} & -0.053 & \textbf{0.232} & \textbf{0.480} & \textbf{0.290} & \textbf{0.359} & \textbf{0.421} & \textbf{0.264} & \textbf{0.669} & \textbf{0.127} & \textbf{0.608} & \textbf{0.637} & \textbf{0.394} & \textbf{0.394} \\
    & PSNR & \textbf{27.77} & \textbf{43.30} & \textbf{38.19} & \textbf{32.94} & \textbf{37.38} & \textbf{40.62} & \textbf{32.15} & \textbf{32.24} & \textbf{38.65} & \textbf{38.01} & \textbf{34.62} & \textbf{32.95} & \textbf{31.82} & \textbf{32.97} & \textbf{36.76} & \textbf{34.39} & \textbf{30.61} & \textbf{35.72} & \textbf{35.06} \\
    \toprule   
    \specialrule{0.1pt}{0pt}{0pt}
    \end{tabular}
    }
    \caption{
    Quantitative comparison on the Orion-CRC dataset
    (\textbf{Bold}: best, \underline{Underline}: second best).
    }
    \label{table:orion}
\end{table*}

\begin{figure*}[t!]
  \centering
   \includegraphics[width=0.92\linewidth]{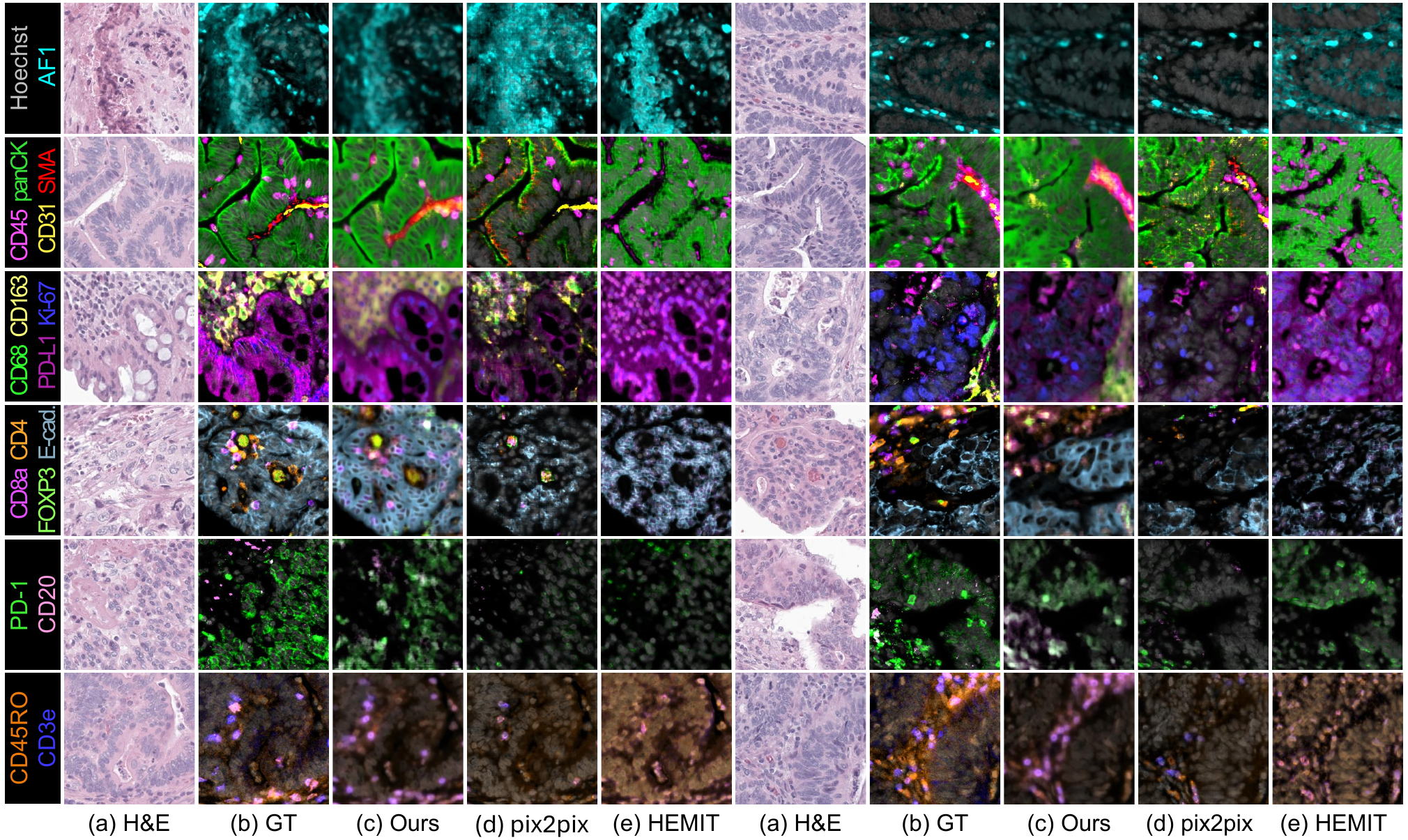}
   \caption{
    Qualitative comparison on Orion-CRC.
    Each row depicts different IF markers in various colors.
    Columns show (a) H\&E, (b) ground truth IF (GT), and (c-e) virtual IF by Ours, pix2pix, and HEMIT, respectively.
   }
   \label{fig:qual}
\end{figure*}

\subsubsection{Baselines.}
We compare our approach to several state-of-the-art methods, including pix2pix~\cite{isola2017image}, pix2pixHD~\cite{wang2018high}, HEMIT~\cite{bian2024hemit},~\citet{parmar2024one}, and Marigold~\cite{ke2024repurposing}.
For the HEMIT dataset with 3 output channels, we implemented Marigold by injecting 3-channel mIHC images into the SD VAE encoder.
pix2pixHD and Marigold were excluded from Orion-CRC evaluations due to architectural limitations for multi-channel (18-marker) generation.
Pix2pix and HEMIT were adapted to output 18 marker channels for the Orion-CRC dataset. 
We retrained HEMIT on its own dataset after observing performance discrepancies with the authors' released checkpoint.
Parmar et al. was excluded from Orion-CRC evaluations due to severe mode collapse. 

\subsection{Results}
\label{subsec:results}
\subsubsection{Performance comparison.} 
In Table~\ref{table:hemit}, on the HEMIT dataset, our approach achieves the highest average SSIM (+0.066), R (+0.040), and PSNR (+1.238) compared to the second-highest scores, achieving the highest in every marker and metric.
Figure~\ref{fig:qual_hemit} further highlights the advantages of our approach, particularly in capturing CD3 marker signals; for visualization, each marker type is shown in a distinct color.
(h) Marigold often fails to localize panCK signals (row 3) and CD3 signals (yellow-box regions), while (g) Parmar et al. tend to over-predict CD3 signals (row 1) or fail to detect CD3 signals (row 3).
(f) HEMIT shows better results than other methods, but our method consistently provides clearer and more accurate localizations, aligning with the quantitative results in Table~\ref{table:hemit}.

On Orion-CRC, Table~\ref{table:orion} demonstrates that our approach achieved the highest average scores, with improvements of +0.039 SSIM, +0.117 R, and +1.358 PSNR compared to the second-best results, achieving the highest SSIM for 13, R for 15, and PSNR for 18 out of 18 markers.
For markers with sufficient signal presence, such as Hoechst (cellular structures), all methods performed well. HEMIT achieved the second-highest performance on Hoechst, but generally underperformed compared to pix2pix on other marker types, with pix2pix consistently showing the second-best performance overall.
Our method showed lower R scores for CD31 and FOXP3, likely a consequence of the limited number of available patches for these markers ($<$1,000). 
Nevertheless, as shown in Fig.~\ref{fig:qual}, our model does not completely miss these signals and still provides reasonable localization (CD31 in row 2, FOXP3 in row 4); in contrast, other methods frequently miss signals for certain marker types such as SMA (row 2) and CD20 (row 5).
While our results exhibit some mild blurriness and tend to predict certain markers over broader regions compared to the ground truth (\textit{e.g.}, panCK in row 2, CD68 and CD163 in row 3), this tendency is likely due to structural properties of the latent diffusion model, in particular the absence of skip connections and its operation in the latent space, as previously observed in~\cite{parmar2024one}.
For the remaining marker types, our method generally matches or exceeds the performance of previous approaches, yielding consistent improvements in both localization accuracy and overall metric scores.

Overall, our method achieves superior localization and performance compared to previous methods, highlighting its potential for H\&E staining analysis supported by virtual multiplex staining.

\begin{table}[t!]
    \setlength{\tabcolsep}{1.2mm}
    \centering
    \begin{tabular}{lllccc}
    \toprule
    \specialrule{0.1pt}{0pt}{0pt}
    Dataset & & Condition & SSIM & R & PSNR (dB) \\
    \hline
    \multirow{2}{*}{HEMIT}
        & & Text     & 0.667 & \textbf{0.772} & 30.09 \\
        & \checkmark & One-hot & \textbf{0.673} & 0.770 & \textbf{30.21} \\
    \hline
    \multirow{2}{*}{Orion-CRC} 
        & & Text     & 0.288 & -0.003 & 18.01 \\
        & \checkmark & One-hot & \textbf{0.662} & \textbf{0.371} & \textbf{30.66} \\
     \toprule
     \specialrule{0.1pt}{0pt}{0pt}
    \end{tabular}
    \caption{Ablation on marker type conditioning strategy}
    \label{tab:ablation_cond}
\end{table}

\begin{figure}[t!]
    \centering
    \includegraphics[width=0.8\linewidth]{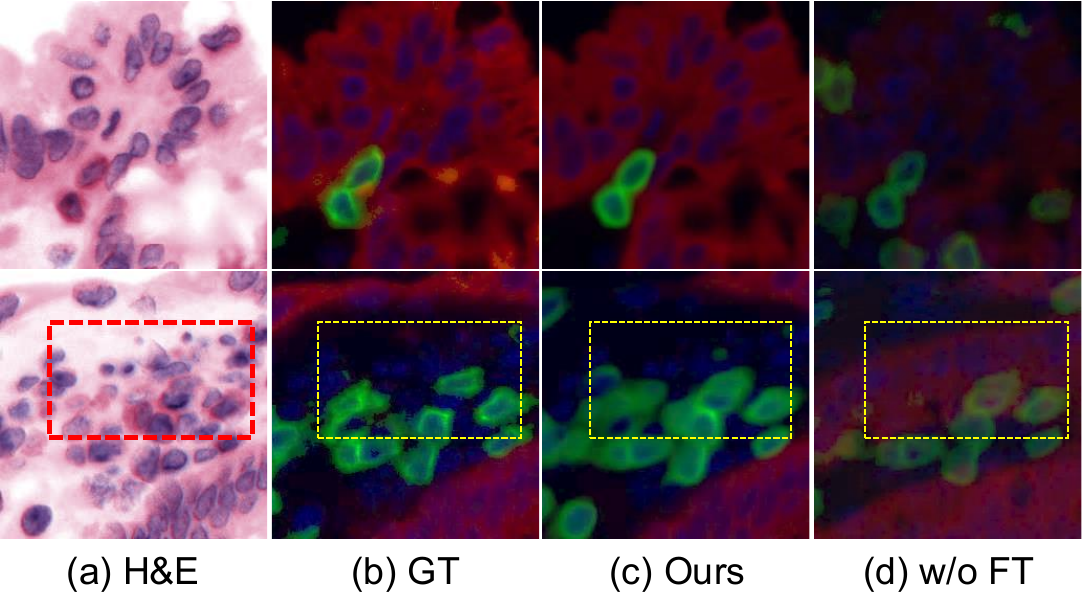}
    \caption{
    Color fidelity comparison on the HEMIT dataset: (a) H\&E, (b) ground truth, (c) Ours, (d) without fine-tuning.
    }
    \label{fig:ablation_color}
\end{figure}

\subsubsection{Ablation studies.}
Table~\ref{tab:ablation_cond} highlights the scalability and effectiveness of marker-wise one-hot embedding approach compared to text-based conditioning, with these results obtained prior to the fine-tuning stage.
On the HEMIT dataset (3 marker types), both strategies produced comparable results, with one-hot conditioning yielding a slightly higher SSIM and PSNR.
However, on Orion-CRC (18 marker types), one-hot conditioning achieved a substantial gain (SSIM 0.662, PSNR 30.663), whereas text conditioning failed to generate outputs aligned with the specified marker types, resulting in much lower SSIM (0.288) and PSNR (18.014).
This confirms that while text-based conditioning can work in small-marker scenarios (like HEMIT),
it does not scale to practical multiplex settings with numerous marker types.
In summary, the ablation demonstrates that marker-wise one-hot embedding is a robust and scalable conditioning strategy for virtual multiplex staining.

\begin{table}[t!]
    \setlength{\tabcolsep}{1.0mm}
    \centering
    {\fontsize{9pt}{9.5pt}\selectfont
    \begin{tabular}{lccccc}
    \toprule
    \specialrule{0.1pt}{0pt}{0pt}
    Method & \makecell{Run time \\ (sec/sample)} & \makecell{Memory \\ (MB)} & SSIM & R & \makecell{PSNR \\ (dB)}  \\
    \midrule
    \multicolumn{6}{l}{(DDIM steps: 50, Ensemble: 10~\cite{ke2024repurposing})} \\ 
    Single-plex 
    & 117.18 & 14607 & 0.670 & 0.755 & \underline{30.32} \\
    Multiplex 
    & 110.59 & 33838 & 0.673 & \underline{0.770} & 30.21 \\
    \midrule
    \multicolumn{6}{l}{(DDIM steps: 1, Ensemble: 1)} \\
    Multiplex 
    & 0.13 & 7850 & \underline{0.757} & 0.760 & 29.38 \\
    +Fine-tune 
    & 0.13 & 7850 & \textbf{0.836} & \textbf{0.795} & \textbf{30.60} \\
    \toprule
    \specialrule{0.1pt}{0pt}{0pt}
    \end{tabular}
    }
    \caption{
    Ablation on the performance and inference cost on the HEMIT dataset.
    }
    \label{tab:ablation_cost}
\end{table}

Figure~\ref{fig:ablation_color} visually compares our fine-tuned model (c) and the non-finetuned model (d) w/o FT.
The model without fine-tuning (d) exhibits significant color distortion and struggles to accurately reproduce the marker-specific colors in the ground truth (b) GT. Additionally, it shows poor localization accuracy, particularly evident in row 2, with false positive panCK (red) predictions in the yellow box.
In contrast, our fine-tuned model (c) achieves superior color fidelity that closely matches the ground truth and demonstrates highly accurate spatial localization of marker signals. This comparison emphasizes the importance of incorporating pixel-level loss-based fine-tuning to enhance both color fidelity and overall model performance.

Table~\ref{tab:ablation_cost} summarizes the impact of the fine-tuning stage on model performance and inference cost on the HEMIT dataset.
A single-plex model (one model per marker) serves as baseline but is impractical for virtual multiplex staining with a large number of markers, requiring multiple models.
Our multiplex model generates all markers in a single network, improving training efficiency and practical scalability.
We used 10-fold test-time ensembling~\cite{ke2024repurposing} for optimal performance.
Multiplex model shows slightly higher SSIM and R, but with much heavier memory demand.
Notably, single-step inference in the multiplex model yields even higher SSIM (0.757) than 50-step sampling with ensembling (0.673).
After the fine-tuning stage, our approach achieves the highest performance with dramatically reduced runtime and memory, demonstrating that fine-tuning is crucial for both image fidelity and efficient, scalable virtual multiplex staining.

\begin{table}[t!]
    \setlength{\tabcolsep}{2.0mm}
    \centering
    {\fontsize{9pt}{9.5pt}\selectfont
    \begin{tabular}{lrcccc}
    \toprule
    \specialrule{0.1pt}{0pt}{0pt}
    Dataset & & $\lambda$ & SSIM & R & PSNR (dB) \\
    \hline
    \multirow{3}{*}{HEMIT} 
        &  & 0 & \textbf{0.837} & 0.788 & \underline{30.71} \\
        & \checkmark & 0.5 & \underline{0.836} & \underline{0.795} & 30.60 \\
        & & 1.0 & 0.803 & \textbf{0.812} & \textbf{31.26} \\
    \hline
    \multirow{3}{*}{Orion-CRC}
        & & 0 & 0.726 & 0.255 & 34.39 \\
        & & 0.5 & 0.741 & 0.283 & 34.58 \\
        & \checkmark & 1.0 & \textbf{0.763} & \textbf{0.394} & \textbf{35.06} \\
     \toprule
     \specialrule{0.1pt}{0pt}{0pt}
    \end{tabular}
    }
    \caption{Ablation on loss weight $\lambda$ in fine-tuning stage}
    \label{tab:ablation_loss}
\end{table}

Table~\ref{tab:ablation_loss} summarizes the effect of the fine-tuning loss weight $\lambda$. 
For HEMIT, we selected $\lambda=0.5$, prioritizing SSIM and R. Notably, our method outperformed all related works in Table~\ref{table:hemit} for any tested $\lambda$.
For Orion-CRC, $\lambda=1.0$ was chosen as it achieved the best overall metric scores.
This variation in optimal $\lambda$, potentially reflecting differences in marker signal distribution between datasets, suggests that dataset adaptive loss function selection may be beneficial for maximizing performance in multiplex settings.

\subsubsection{Discussion.}
While our framework achieves superior performance for virtual multiplex staining, several challenges remain.
%
Single-step inference reduces per-sample cost, but computational cost scales with marker count, which is an inherent challenge for high-dimensional applications.
%
Our model also lacks explicit inter-marker correlation modeling, potentially limiting image fidelity. 
Addressing these points is crucial for practical deployment.

\section{Conclusion}
\label{sec:conclusion}

In this paper, we proposed a novel framework for virtual multiplex staining using marker-wise conditioned diffusion models.
By conditioning on marker-specific embeddings and refining with pixel-level loss functions, we achieved state-of-the-art performance on public datasets while improving practical applicability.
Future directions include enhancing structural detail, explicitly modeling inter-marker correlations, and evaluating clinical impact in pathology.

\section*{Acknowledgements}
This work was supported in part by the National Research Foundation of Korea under Grant RS-2024-00349697 and Grant RS-2021-NR060143; in part by the Institute for Information and Communications Technology Planning and Evaluation under Grant IITP-2025-RS-2020-II201819; in part by the Technology Development Program funded by the Ministry of SMEs and Startups (MSS), South Korea, under Grant RS-2024-00437796; in part by the National Research Council of Science and Technology (NST) grant funded by Korean Government [Ministry of Science and Information and Communications Technology (MSIT)] under Grant GTL24031-000; and in part by Korea University Grant.


\bibliography{aaai2026}


\clearpage
\FloatBarrier
\begin{center}
    \addcontentsline{toc}{chapter}{Supplementary Material}
    \LARGE \textbf{Supplementary Material}
\end{center}

\begin{figure}[t]
    \centering
    \includegraphics[width=0.95\linewidth]{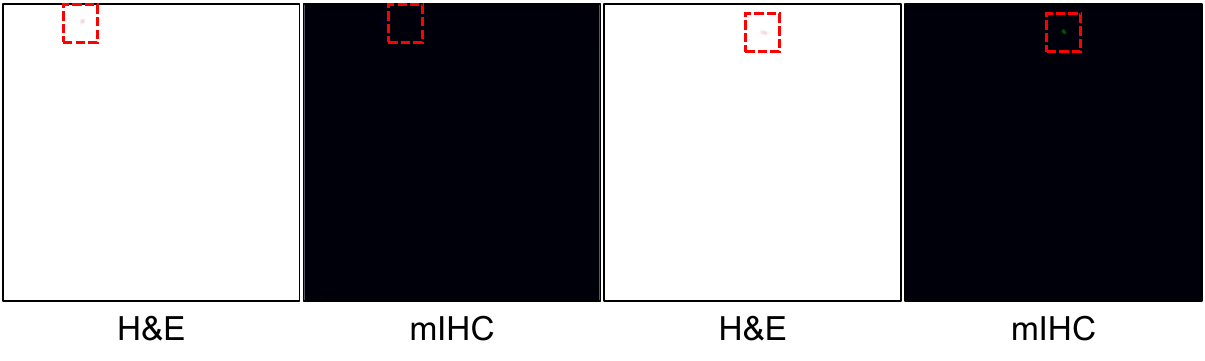}
    \caption{
    Examples of excluded H\&E and mIHC patch pairs from the test set of the HEMIT dataset. These biologically irrelevant, low-signal regions can artificially inflate evaluation metrics such as PSNR (e.g., $>$60). Each patch visualized is 1,024$\times$1,024 pixels in size.
    }
    \label{fig:hemit_excluded}
\end{figure}

\section{More Details on Datasets}
In this section, we provide additional details on the datasets.

\begin{figure}[t]
    \centering
    \includegraphics[width=\linewidth]{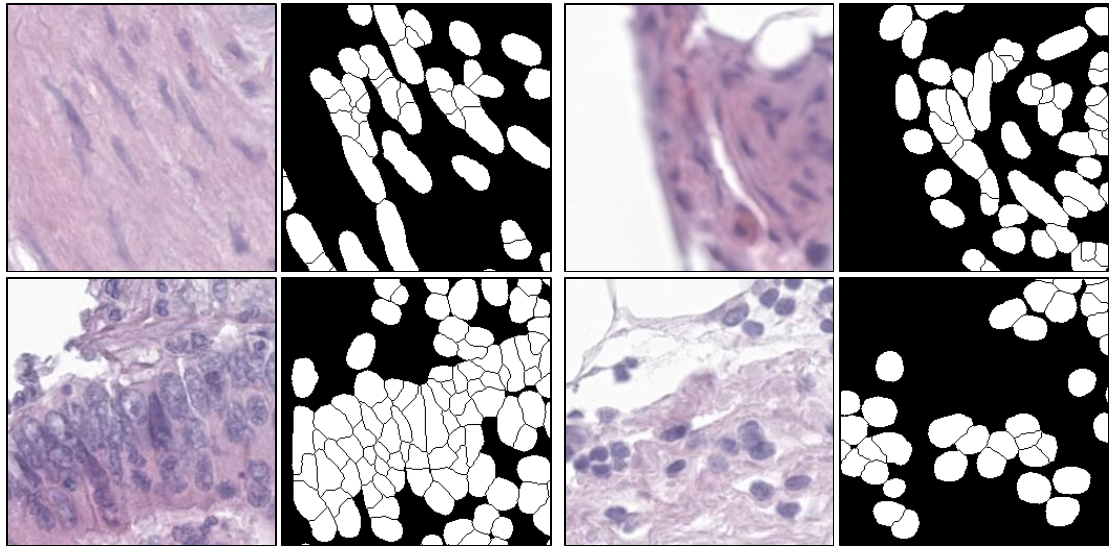}
    \caption{
    Visualization of H\&E tissue image patches from the Orion-CRC dataset and their corresponding binary cell segmentation masks~\cite{lin2023high}. Each patch visualized is 256$\times$256 pixels in size. The segmentation masks were used to filter out patches with low cell content for experimental analysis.
    }
    \label{fig:seg_mask}
\end{figure}

\subsection{Setup on HEMIT}
Figure~\ref{fig:hemit_excluded} illustrates examples of `empty' patches that we excluded from quantitative evaluation. Such empty or low-signal regions patches can artificially inflate metric scores, with PSNR values often exceeding 60, and thus hinder meaningful quantitative assessment.
To ensure reliable evaluation, we excluded these patches from our analysis.


\subsection{Setup on Orion-CRC} 

\subsubsection{Patch filtering.} 
Figure~\ref{fig:seg_mask} shows example H\&E tissue images from the Orion-CRC~\cite{lin2023high} dataset alongside their paired binary cell segmentation masks, each visualized at a size of 256$\times$256 pixels.
During dataset preparation, these binary masks were used to filter patches. As a result, only patches with at least 25\% of the area covered by cells were included in the experiment.

\subsubsection{Quantitative evaluation.} Table~\ref{tab:num_test_orion} summarizes the distribution of patches for different immunofluorescence (IF) markers in the Orion-CRC dataset used for quantitative evaluation. The number of patches per marker was determined after applying an SSIM thresholding. Each patch's SSIM value was calculated using an empty array as the reference, and only patches with an SSIM $\leq0.8$ to the empty array were retained.

\begin{table}[t]
\small
\setlength{\tabcolsep}{1.3mm}
\centering
\begin{tabular}{lr|lr|lr}
\toprule
\specialrule{0.1pt}{0pt}{0pt}
\multicolumn{1}{l}{\textbf{Marker}} & \multicolumn{5}{r}{\# of patches used for quantitative analysis} \\
\hline
\textbf{Hoechst} & 39,625 & \textbf{FOXP3} & 4 & \textbf{CD163} & 3,438 \\
\textbf{AF1} & 1,116 & \textbf{CD8a} & 911 & \textbf{E-cadherin} & 21,285 \\ 
\textbf{CD31} & 23 & \textbf{CD45RO} & 22,505 & \textbf{PD-1} & 3,480 \\
\textbf{CD45} & 25,688 & \textbf{CD20} & 2,214 & \textbf{Ki67} & 5,194 \\
\textbf{CD68} & 1,235 & \textbf{PD-L1} & 18,494 & \textbf{panCK} & 22,361 \\
\textbf{CD4} & 1,022 & \textbf{CD3e} & 16,985 & \textbf{SMA} & 4,356 \\
\toprule
\specialrule{0.1pt}{0pt}{0pt}
\end{tabular}
\caption{Number of patches per IF marker in the Orion-CRC~\cite{lin2023high} dataset used for quantitative analysis.} 
\label{tab:num_test_orion}
\end{table}


\begin{figure*}
    \centering
    \includegraphics[width=1.0\linewidth]{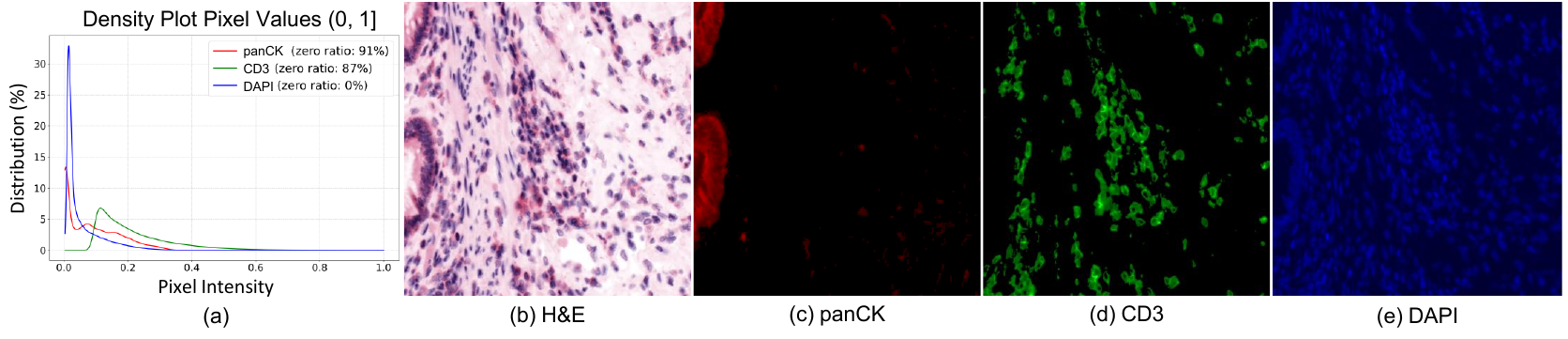}
    \caption{Comparison of pixel intensity distributions for different markers in an example image from the HEMIT~\cite{bian2024hemit} dataset.
    (a) Density plots of normalized pixel intensities, showing the distribution of pixel values in the range (0, 1] and the corresponding zero-ratios for each mIHC marker channel.
    (b) H\&E image, (c) panCK (zero-ratio: 91\%), (d) CD3 (zero-ratio: 87\%), and (e) DAPI (zero-ratio: 0\%).
    }
    \label{fig:color_dist_hemit}
\end{figure*}

\subsection{Marker Type Distributions}
Fig.~\ref{fig:color_dist_hemit} shows the pixel intensity characteristics of different markers in an example image from the HEMIT dataset~\cite{bian2024hemit}.
Density plot (a) quantitatively illustrates the percentage distribution of normalized pixel values (range: 0-1, excluding 0) and the proportion of background pixels (zero-ratio) for each marker channel of the example image.
The example consists of an H\&E image (b) paired with the corresponding mIHC marker images (c-e).
The analysis reveals significant differences in marker expression patterns and background prevalence across markers.
For clarity of visualization, the contrast and brightness of mIHC images (c-e) have been adjusted.

\section{Further Implementation Details}
\subsection{Marker-wise Conditional Diffusion Model}
\label{sec:sub_cdm}
We followed the implementation settings of~\citet{ke2024repurposing} with an effective batch size of 32. 
The Adam optimizer was employed with an initial learning rate of 3e-5, a 100-step warm-up period, and exponential decay over 30,000 iterations.
All experiments used four NVIDIA H100 GPUs.
The DDPM noise scheduler~\cite{ho2020denoising} was set with $T=1000$ timesteps and incorporated multi-resolution noise~\cite{whitaker2024multi}.
Text conditioning was disabled by injecting an empty token.

\subsubsection{HEMIT:} We used 8 images per GPU, totaling 96 images per update step (8 images $\times$ 4 GPUs $\times$ 3 markers).

\subsubsection{Orion-CRC:} Batch size was set to 2 per GPU, with gradient accumulation over 4 steps, resulting in an effective 576 images per update (2 images $\times$ 4 GPUs $\times$ 4 steps $\times$ 18 markers). 
%

\subsection{Fine-tuning for Color Contrast Fidelity}
The fine-tuning stage used the same learning rate schedule as above, starting at an initial learning rate of 3e-5 with a 100-step warm-up and exponential decay.
Due to increased memory requirements for pixel-level loss backpropagation, batch sizes were reduced compared to~\ref{sec:sub_cdm}.

\subsubsection{HEMIT:} We set the batch size to 2, accumulating gradients over 2 steps on 4 GPUs, yielding 48 images per update  step (2 $\times$ 2 $\times$ 4 $\times$ 3), with H\&E images replicated across three markers. Fine-tuning ran for 5,000 iterations over 9 hours. After 5,000 iterations, performance improvements saturated.

\subsubsection{Orion-CRC:} Batch size was 6, with accumulation over 2 steps and 4 GPUs, processing 48 images per update step without H\&E image replication due to memory constraints. Fine-tuning ran for 50,000 iterations (about 2.5 days). Performance improvements became marginal after 30,000 iterations.
To improve efficiency on the Orion-CRC dataset, We randomly sub-sampled 20\% of markers in a marker-wise fashion at this stage.


\section{Marker-wise Results on Ablation Study}
In this section, we provide comprehensive per-marker results for our ablation studies, which were omitted from the main paper due to space constraints. The detailed tables present the performance metrics for each marker.

\begin{table*}[!ht]
\centering
\setlength{\tabcolsep}{2.0mm}
{
\begin{tabular}{lcccccccccccc}
\toprule   
\specialrule{0.1pt}{0pt}{0pt}
& \multicolumn{4}{c}{SSIM} & \multicolumn{4}{c}{R} & \multicolumn{4}{c}{PSNR (dB)} \\
\cmidrule(lr){2-5} \cmidrule(lr){6-9} \cmidrule(lr){10-13}
Method & DAPI & CD3 & panCK & Avg. & DAPI & CD3 & panCK & Avg. & DAPI & CD3 & panCK & Avg. \\

\midrule

Text
& 0.837 & 0.616 & 0.547 & 0.667 & 0.968 & 0.578 & 0.770 & 0.772 & 35.152 & 26.052 & 29.079 & 30.094 \\

One-hot
& 0.838 & 0.634 & 0.548 & 0.673 & 0.967 & 0.569 & 0.775 & 0.770 & 35.380 & 25.978 & 29.275 & 30.211 \\

\toprule   
\specialrule{0.1pt}{0pt}{0pt}
\end{tabular}
}
\caption{
Ablation study on marker type conditioning strategy, reporting per-marker metric scores on the HEMIT dataset.
}
\label{table:hemit_cond_detailed}
\end{table*}

\begin{table*}[t!]
    \setlength{\tabcolsep}{0.3mm}
    {\fontsize{9pt}{9.5pt}\selectfont 
    \centering
    \begin{tabular}{c| l c c c c c c c c c c c c c c c c c c c}
    \toprule
    \specialrule{0.1pt}{0pt}{0pt} 
    \rotatebox{90}{Method} & \; \ \rotatebox{90}{Metric} & \rotatebox{90}{Hoechst} & \rotatebox{90}{AF1} & \rotatebox{90}{CD31} & \rotatebox{90}{CD45} & \rotatebox{90}{CD68} & \rotatebox{90}{CD4} & \rotatebox{90}{FOXP3} & \rotatebox{90}{CD8a} & \rotatebox{90}{CD45RO} & \rotatebox{90}{CD20} & \rotatebox{90}{PD-L1} & \rotatebox{90}{CD3e} & \rotatebox{90}{CD163} & \rotatebox{90}{E-cad.} & \rotatebox{90}{PD-1} & \rotatebox{90}{Ki67} & \rotatebox{90}{panCK} & \rotatebox{90}{SMA} & \rotatebox{90}{Avg.} \\
    \midrule
    \multirow{3}{*}{\makecell{\rotatebox{90}{Text}}} 
    & SSIM & 0.130 & 0.327 & 0.322 & 0.256 & 0.331 & 0.324 & 0.307 & 0.315 & 0.278 & 0.289 & 0.274 & 0.300 & 0.302 & 0.257 & 0.325 & 0.294 & 0.247 & 0.301 & 0.288 \\
    & R & -0.002 & -0.006 & -0.006 & -0.002 & -0.005 & -0.006 & -0.006 & -0.002 & -0.002 & -0.004 & -0.001 & -0.002 & -0.004 & -0.002 & -0.003 & -0.002 & -0.001 & -0.003 & -0.003 \\
    & PSNR & 15.84 & 17.98 & 17.87 & 17.78 & 18.20 & 18.16 & 18.33 & 18.08 & 18.32 & 18.09 & 18.41 & 18.14 & 18.18 & 18.56 & 18.30 & 17.98 & 17.99 & 18.05 & 18.01 \\
    
    \midrule
    
    \multirow{3}{*}{\makecell{\rotatebox{90}{One-hot}}} 
    & SSIM & 0.626 & 0.808 & 0.792 & 0.640 & 0.759 & 0.686 & 0.768 & 0.615 & 0.768 & 0.734 & 0.660 & 0.648 & 0.610 & 0.666 & 0.726 & 0.728 & 0.625 & 0.638 & 0.694 \\
    &  R   & 0.765 & 0.300 & 0.029 & 0.301 & 0.126 & 0.155 & 0.031 & 0.052 & 0.248 & 0.134 & 0.152 & 0.181 & 0.059 & 0.404 & 0.097 & 0.281 &  0.394 & 0.136 & 0.214  \\
    & PSNR & 24.19 & 40.74 & 39.04 & 30.97 & 36.97 & 36.93 & 33.65 & 31.51 & 35.79 & 36.93 & 32.72 & 31.62 & 31.33 & 31.33 & 36.80 & 33.61 & 29.36 & 33.97 & 33.75 \\
    \toprule   
    \specialrule{0.1pt}{0pt}{0pt}
    \end{tabular}
    }
    \caption{
    Ablation study on marker type conditioning strategy, reporting per-marker metric scores on the Orion-CRC dataset. Evaluation was performed using models prior to the fine-tuning stage.
    }
    \label{table:orion_cond_detailed}
\end{table*}

\begin{table*}[!ht]
\centering
\setlength{\tabcolsep}{2.0mm}
{
\begin{tabular}{lcccccccccccc}
\toprule   
\specialrule{0.1pt}{0pt}{0pt}
& \multicolumn{4}{c}{SSIM} & \multicolumn{4}{c}{R} & \multicolumn{4}{c}{PSNR (dB)} \\
\cmidrule(lr){2-5} \cmidrule(lr){6-9} \cmidrule(lr){10-13}
Method & DAPI & CD3 & panCK & Avg. & DAPI & CD3 & panCK & Avg. & DAPI & CD3 & panCK & Avg. \\

\midrule

$\lambda=0$
& \underline{0.860} & \textbf{0.890} & \underline{0.761} & \textbf{0.837} & \textbf{0.973} & 0.612 & 0.777 & 0.788 & 35.986 & 26.313 & 29.838 & \underline{30.713} \\

$\lambda=0.5$
& 0.855 & \underline{0.889} & \textbf{0.763} & \underline{0.836} & \underline{0.972} & \underline{0.633} & \underline{0.781} & \underline{0.795} & 35.398 & \underline{26.527} & \underline{29.858} & 30.595 \\

$\lambda=1.0$
& \textbf{0.868} & \textbf{0.788} & 0.752 & 0.803 & \underline{0.972} & \textbf{0.676} & \textbf{0.789} & \textbf{0.812} & \textbf{37.166} & \textbf{26.696} & \textbf{29.905} & \textbf{31.256} \\
\toprule   
\specialrule{0.1pt}{0pt}{0pt}
\end{tabular}
}
\caption{
Ablation study of loss weight $\lambda$ during fine-tuning stage, reporting per-marker metric scores on the HEMIT dataset (\textbf{Bold}: Best, \underline{Underline}: Second best).
}
\label{table:hemit_lambda_detailed}
\end{table*}

\begin{table*}[t!]
    \setlength{\tabcolsep}{0.68mm}
    {\fontsize{9pt}{9.5pt}\selectfont 
    \centering
    \begin{tabular}{c| l c c c c c c c c c c c c c c c c c c c}
    \toprule
    \specialrule{0.1pt}{0pt}{0pt} 
    \rotatebox{90}{Method} & \; \ \rotatebox{90}{Metric} & \rotatebox{90}{Hoechst} & \rotatebox{90}{AF1} & \rotatebox{90}{CD31} & \rotatebox{90}{CD45} & \rotatebox{90}{CD68} & \rotatebox{90}{CD4} & \rotatebox{90}{FOXP3} & \rotatebox{90}{CD8a} & \rotatebox{90}{CD45RO} & \rotatebox{90}{CD20} & \rotatebox{90}{PD-L1} & \rotatebox{90}{CD3e} & \rotatebox{90}{CD163} & \rotatebox{90}{E-cad.} & \rotatebox{90}{PD-1} & \rotatebox{90}{Ki67} & \rotatebox{90}{panCK} & \rotatebox{90}{SMA} & \rotatebox{90}{Avg.} \\
    \midrule
    \multirow{3}{*}{\makecell{\rotatebox{90}{$\lambda=0$}}} 
    & SSIM & \underline{0.787} & 0.832 & 0.764 & 0.765 & 0.728 & 0.658 & \underline{0.623} & \underline{0.713} & 0.835 & 0.580 & \textbf{0.755} & 0.720 & \underline{0.722} & \textbf{0.772} & 0.699 & 0.660 & \underline{0.763} & 0.684 & 0.726 \\
    & R & \underline{0.910} & \underline{0.549} & -0.006 & 0.544 & -0.025 & 0.067 & \textbf{-0.053} & \underline{0.034} & \underline{0.472} & \underline{0.022} & 0.320 & 0.346 & -0.016 & 0.657 & 0.035 & 0.124 & 0.629 & -0.019 & 0.255 \\
    & PSNR & 27.70 & 42.87 & 37.96 & 32.49 & 36.58 & 38.74 & \underline{32.13} & 31.43 & \textbf{38.68} & 36.76 & \textbf{34.65} & 32.23 & 31.06 & \underline{32.76} & 36.14 & 32.45 & 30.33 & 34.03 & 34.39 \\
    \midrule
    \multirow{3}{*}{\makecell{\rotatebox{90}{$\lambda=0.5$}}} 
    & SSIM & \textbf{0.788} & \textbf{0.857} & \underline{0.769} & \textbf{0.773} & \underline{0.754} & \underline{0.764} & \textbf{0.625} & \textbf{0.717} & \textbf{0.840} & \underline{0.591} & \underline{0.753} & \textbf{0.733} & \textbf{0.725} & 0.770 & \underline{0.758} & \underline{0.669} & \textbf{0.771} & \underline{0.687} & \underline{0.741} \\
    & R & \textbf{0.912} & \textbf{0.560} & \underline{-0.005} & \underline{0.557} & \underline{0.019} & \underline{0.222} & \underline{-0.079} & 0.033 & 0.468 & 0.020 & \underline{0.329} & \underline{0.393} & \underline{-0.011} & \textbf{0.671} & \textbf{0.138} & \underline{0.227} & \textbf{0.642} & \underline{0.000} & \underline{0.283} \\
    & PSNR & \textbf{27.85} & \textbf{43.44} & \underline{38.02} & \underline{32.68} & \underline{36.77} & \underline{39.98} & \underline{32.13} & \underline{31.45} & \underline{38.67} & \underline{36.83} & 34.57 & \underline{32.49} & \underline{31.08} & 32.68 & \underline{36.73} & \underline{32.50} & \underline{30.49} & \underline{34.05} & \underline{34.58} \\
    \midrule
    \multirow{3}{*}{\textbf{\rotatebox{90}{$\lambda=1.0$}}}
    & SSIM & 0.779 & \underline{0.852} & \textbf{0.785} & \underline{0.770} & \textbf{0.815} & \textbf{0.803} & \underline{0.623} & 0.712 & \underline{0.838} & \textbf{0.718} & \underline{0.753} & \underline{0.724} & 0.718 & \underline{0.771} & \textbf{0.764} & \textbf{0.793} & 0.747 & \textbf{0.761} & \textbf{0.763} \\
    & R   & \textbf{0.912} & {0.510} & \textbf{0.027} & \textbf{0.563} & \textbf{0.319} & \textbf{0.340} & \textbf{-0.053} & \textbf{0.232} & \textbf{0.480} & \textbf{0.290} & \textbf{0.359} & \textbf{0.421} & \textbf{0.264} & \underline{0.669} & \underline{0.127} & \textbf{0.608} & \underline{0.637} & \textbf{0.394} & \textbf{0.394} \\
    & PSNR & \underline{27.77} & \underline{43.30} & \textbf{38.19} & \textbf{32.94} & \textbf{37.38} & \textbf{40.62} & \textbf{32.15} & \textbf{32.24} & {38.65} & \textbf{38.01} & \underline{34.62} & \textbf{32.95} & \textbf{31.82} & \textbf{32.97} & \textbf{36.76} & \textbf{34.39} & \textbf{30.61} & \textbf{35.72} & \textbf{35.06} \\
    \toprule   
    \specialrule{0.1pt}{0pt}{0pt}
    \end{tabular}
    }
    \caption{
    Ablation study of loss weight $\lambda$ during fine-tuning stage, reporting per-marker metric scores on the Orion-CRC dataset
    (\textbf{Bold}: best, \underline{Underline}: second best).
    }
    \label{table:orion_lambda_detailed}
\end{table*}

\subsection{Conditioning Strategy}
Table~\ref{table:hemit_cond_detailed} and Table~\ref{table:orion_cond_detailed} provide marker-wise metric scores for the ablation study of the marker type conditioning strategy on the HEMIT and Orion-CRC datasets, respectively, comparing text-based and one-hot conditioning before the fine-tuning stage.

\subsection{Hyperparameter $\lambda$ for Fine-tuning Stage}
Table~\ref{table:hemit_lambda_detailed} and Table~\ref{table:orion_lambda_detailed} present per-marker metric scores for the ablation study of the loss weight $\lambda$ on the HEMIT and Orion-CRC datasets, respectively.







\end{document}